Scanning tunneling microscopy study of helimagnetic monolayer CrBr$_2$ on $s$-wave superconductor NbSe$_2$: a topologically trivial system due to weak interfacial coupling

Yuanji Li[1], Ruotong Yin[1], Mingzhe Li[1], Shiyuan Wang[1], Jiashuo Gong[1], Ziyuan Chen[1], Jiakang Zhang[1], Dong-Lai Feng[2], Ya-Jun Yan[1,2*]

[1]Hefei National Research Center for Physical Sciences at the Microscale and Department of Physics, University of Science and Technology of China, Hefei, 230026, China
[2]New Cornerstone Science Laboratory, Hefei National Laboratory, Hefei 230088, China

*Corresponding author: yanyj87@ustc.edu.cn

**Hybrid magnet-superconductor heterostructures attract significant interest for their potential to host unconventional superconductivity, topological superconductivity, and Majorana physics. Transition metal dihalides (MX$_2$, M = transition metal, X = Cl, Br, I) are compelling magnetic candidates due to their novel magnetic structures and possible ferroelectricity. Here, we employ low-temperature scanning tunneling microscopy/spectroscopy to investigate the interfaces fabricated by growing helimagnet candidate CrBr$_2$ on $s$-wave superconductor NbSe$_2$. Our results reveal that the monolayer CrBr$_2$ is insulating, the measured low-energy electronic states on it derive entirely from the NbSe$_2$ substrate. The superconducting properties of CrBr$_2$/NbSe$_2$ are nearly identical to the bare NbSe$_2$, manifested by the superconducting gap spectra and their temperature and magnetic field dependence, as well as the spatial distribution and bound states of magnetic vortices. Furthermore, in-gap excitations appear only at the dirty edges of CrBr$_2$ islands and are absent from clean edges, suggesting the lack of intrinsic edge states. Taken together, these findings establish the topologically trivial nature of the helimagnetic insulator/$s$-wave superconductor system CrBr$_2$/NbSe$_2$, attributable to the absence of interfacial superconducting proximity and weak magnetic coupling.**

## I. Introduction

Majorana zero modes (MZM) are exotic zero-energy bound states inherent in topological superconductors (TSC) [1-3]. Governed by non-Abelian statistics, MZMs in principle can constitute the fundamental basis for fault-tolerant topological quantum computation and have consequently been the subject of widespread research in recent years [4,5]. Since intrinsic TSCs are extremely rare, various schemes have been proposed for fabricating artificial TSCs, such as the Fu-Kane model for topological insulator-superconductor heterostructures [6], hybrid superconductor-semiconductor nanowire devices [7], magnet-superconductor heterostructures [8], etc. Among them, magnet-superconductor heterostructures have gradually emerged as a promising route to realize topological superconductivity and Majorana modes, due to the abundance of available magnetic materials and the simplicity of heterostructure construction [9-22].

Two-dimensional (2D) magnetic materials, a booming field in recent years, exhibit simplified structural configurations, facile growth processes, and robust retention of magnetic properties even in their monolayer form [23], prompting growing research interests in 2D magnetic material-superconductor hybrid systems. Transition metal dihalides, MX$_2$ (M represents transition metals and X represents Cl, Br, or I), are among the simplest types of 2D magnetic materials. Neutron scattering experiments have revealed helical magnetism in CoI$_2$ [24], NiI$_2$ [24], MnI$_2$ [25], MnBr$_2$

[26], CrI$_2$ [27], and CrBr$_2$ [28], as well as a spin spiral structure in CuCl$_2$ [29]. These noncollinear magnetic structures, on one hand, may induce topological superconducting states and Majorana modes when coupled with superconductivity in the form of heterostructures [30-34]; on the other hand, they can spontaneously generate magnetoelectric coupling effects such as ferroelectricity, antiferroelectricity, and multiferroicity within the materials [35-39], which can be further utilized to manipulate the topological superconducting states [40-42]. Therefore, heterostructures of transition metal dihalides and superconductors may be a promising platform for simultaneously generating and tuning topological superconductivity and Majorana modes. However, experimental research in this regard is still very scarce, and the required range of interfacial coupling parameters for realizing topological superconductivity remains largely unexplored.

Here, we grow monolayer CrBr$_2$ film on NbSe$_2$ superconductor (hereafter labeled as CrBr$_2$/NbSe$_2$) and study its properties by using low-temperature scanning tunneling microscopy/spectroscopy (STM/STS). The monolayer CrBr$_2$ film is insulating, thus there is no superconducting proximity on it; instead, it acts as a vacuum tunneling barrier in STM study, facilitating the detection of superconducting gap and vortex states of the underlying NbSe$_2$, which are nearly identical to those of bare NbSe$_2$ substrate. Moreover, a hard superconducting gap without in-gap states is observed at the clean edges of CrBr$_2$ islands, while distinct in-gap states appear at the dirty edges, which are assigned as Yu-Shiba-Rusinov (YSR) states. These findings indicate that the conventional $s$-wave superconducting characteristics of NbSe$_2$ remain essentially unchanged at the CrBr$_2$/NbSe$_2$ interface and topological superconducting states are not induced. By comparing with other magnet-superconductor heterostructure systems and performing an estimation of the interfacial magnetic exchange coupling strength, we conclude that the topologically trivial nature in magnetic insulator-superconductor heterostructures is caused by weak interfacial coupling due to the absence of superconducting proximity and weak magnetic coupling. Based on this, several schemes are suggested to enhance the interfacial coupling in the future.

## II. Methods

CrBr$_2$ films were grown by molecular beam epitaxy (MBE) method in the prepare chamber equipped on an ultralow-temperature STM (UNISOKU 1600). NbSe$_2$ single crystals were grown by chemical vapor transport (CVT) method and show a typical superconducting transition temperature of ~ 7.0 K. CrBr$_3$ single crystals were used as the evaporation sources, which were also grown by CVT method. For film growth, NbSe$_2$ crystals were mechanically cleaved at ~ 80 K in ultrahigh vacuum with a base pressure better than $2 \times 10^{-10}$ mbar; monolayer CrBr$_2$ films were obtained by evaporating CrBr$_3$ molecules from the Knudsen cell (450 °C) to NbSe$_2$ substrate (270-280 °C), and then post-annealed at 270 °C for ~ 5 minutes to improve the film quality. Then the sample was transferred into the STM chamber for STM study. PtIr tips were used for STM study after being treated on a clean Au (111) substrate. The d$I$/d$V$ spectra were collected by a standard lock-in technique with a modulation frequency of 973 Hz. The data in the main text were collected at a temperature of ~ 30 mK with an effective electron temperature $T_{eff}$ of ~ 170 mK [43].

## III. Results

### A. Structural and electronic properties of CrBr$_2$/NbSe$_2$ heterostructure

CrBr$_2$ crystallizes in a layered structure (space group $C/m$) [44], with each layer consisting of a Cr sublattice sandwiched between two Br sublattices [top panel of Fig. 1(a)]. Due to Jahn-Teller distortion [28,44], three adjacent Br atoms in the same sublayer form an isosceles triangle, with the apex angle $\theta \sim 54.3°$ [orange triangle in the bottom panel of Fig. 1(a)]. Figure 1(b) shows the typical topographic image of CrBr$_2$/NbSe$_2$ heterostructures, the CrBr$_2$ film does not cover the entire surface, as evidenced by the dark holes on it and the sharp step edges with NbSe$_2$ substrate. Figure 1(c) displays the height profile along the magenta line (cut #1) as indicated in Fig. 1(b), revealing a step height of approximately 0.45 nm, smaller than the c-axis lattice constant of ~ 0.62 nm for bulk CrBr$_2$. This reduced step height should be attributed to the modified interface induced by the different interlayer stacking patterns and interactions between NbSe$_2$ and CrBr$_2$.

Figure 1(d) displays the detailed topographic image of CrBr$_2$ film, an elongated hexagonal moiré lattice is formed, arising from the in-plane lattice mismatch both in lattice parameter and symmetry between NbSe$_2$ (three adjacent Se atoms form an equilateral triangle with the side length $a = 0.344$ nm) and CrBr$_2$ (side lengths of the isosceles triangle are $a_1 = a_2 = 0.399$ nm and $a_3 = 0.365$ nm). The unit cell of the moiré lattice is illustrated by the yellow rhomboid in Fig. 1(d), exhibiting a period of ~ 3.7 nm × 3.7 nm. Figure 1(e) displays the topographic image of CrBr$_2$ film with much higher resolution, and the inset is an enlarged view. The three adjacent Br atoms do form an isosceles triangle, consistent with the crystal structure of bulk CrBr$_2$.

We further investigate the electronic properties of CrBr$_2$ film, its typical d$I$/d$V$ spectra are shown in Figs. 1(f) and 1(g), with those on bare NbSe$_2$ listed as well for comparison. As shown in Fig. 1(f), the d$I$/d$V$ spectra within ±3 eV on CrBr$_2$/NbSe$_2$ and bare NbSe$_2$ differ significantly, the former shows an insulating band gap between -1.5 eV and 2.7 eV, in contrast to the metallic behavior of bare NbSe$_2$. However, for the measured energy range within the band gap of CrBr$_2$ [Fig. 1(g)], two nearly identical spectra are observed on CrBr$_2$/NbSe$_2$ and bare NbSe$_2$. This is because the CrBr$_2$ film has no density of state (DOS) in the measured energy range of Fig. 1(g) and behaves as a vacuum tunneling barrier, through which electrons tunnel into the underlying NbSe$_2$ layer. This phenomenon is similar to previous studies on ferromagnetic insulator/superconductor heterostructure CrBr$_3$/NbSe$_2$ [45-47], and has also been widely observed by STM studies on insulator/superconductor heterostructures, such as MnTe/NbSe$_2$ [48], CrI$_2$/NbSe$_2$ [49], and CrCl$_3$/NbSe$_2$ [50]. Furthermore, when scrutinizing the spectral features around 250 meV, we find the spectrum on CrBr$_2$/NbSe$_2$ is shifted slightly towards higher energies by ~30 meV compared to that of bare NbSe$_2$, as marked out in the inset of Fig. 1(g). Similar rigid band shifts have been reported previously in CrBr$_3$/NbSe$_2$ and CrCl$_3$/NbSe$_2$ heterostructures [45-47,50], which should be mainly attributed to work function mismatch and weak charge redistribution at the interfaces, as supported by DFT calculations in refs. [45,46].

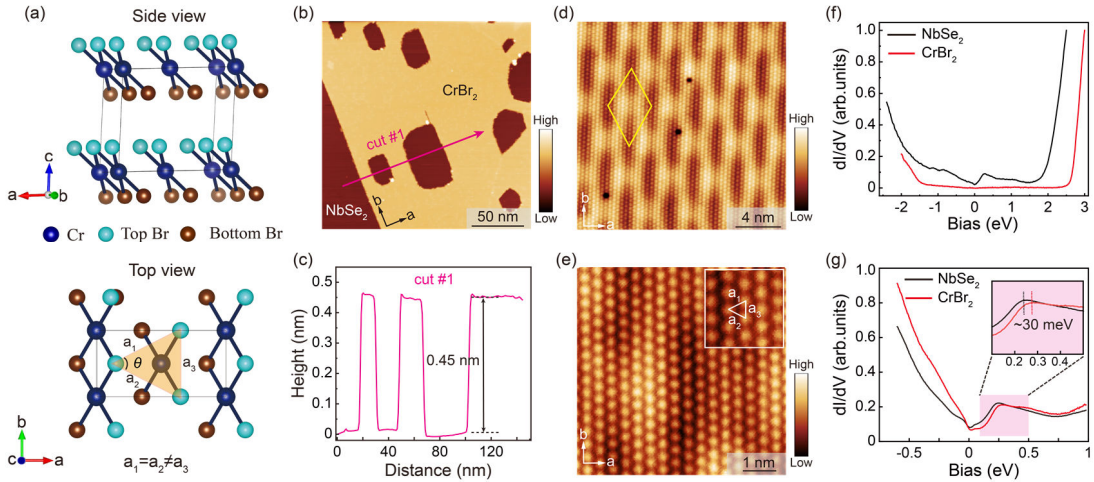

**Fig. 1**. **Structural and electronic properties of CrBr$_2$/NbSe$_2$ heterostructure**. (a) Crystal structure of bulk CrBr$_2$. The three adjacent Br atoms in the same sublayer form an isosceles triangle (orange triangle in the bottom panel).

(b) Typical topographic image of CrBr$_2$/NbSe$_2$ heterostructure. (c) Height profile taken along cut #1 in panel (b). (d) Typical topographic image of CrBr$_2$ film. The unit cell of Moiré superlattice is indicated by the yellow rhombus. (e) Topographic image of CrBr$_2$ film with higher resolution, the inset shows the enlarged view of atom arrangements with $a_1 = a_2 \neq a_3$. (f),(g) Typical d$I$/d$V$ spectra measured on CrBr$_2$/NbSe$_2$ and bare NbSe$_2$ substrate. Inset of panel (g) shows the enlarged view of the shaded region. Measurement conditions: (b) $V_b$ = 3 V, $I_t$ = 20 pA; (d) $V_b$ = 1.5 V, $I_t$ = 100 pA; (e) $V_b$ = 1 V, $I_t$ = 50 pA; (f) $V_b$ = 2.5 V, $I_t$ = 200 pA, $\Delta V$ = 50 mV for NbSe$_2$, and $V_b$ = 3 V, $I_t$ = 200 pA, $\Delta V$ = 50 mV for CrBr$_2$; (g) $V_b$ = 0.98 V, $I_t$ = 200 pA, $\Delta V$ = 20 mV.

## B. Superconducting gap spectra of CrBr$_2$/NbSe$_2$ heterostructure

Figure 2(a) shows the topographic image of a selected sample region of CrBr$_2$/NbSe$_2$ heterostructure, capturing both the heterostructure and the bare NbSe$_2$ substrate, separated by a single-unit-cell-height step. At $T$ = 30 mK, the superconducting gap spectra measured across this step are presented in Fig. 2(b), revealing a fully developed superconducting gap that persists along the entire trajectory. Figure 2(c) compares in more detail the averaged superconducting gap spectra obtained on CrBr$_2$/NbSe$_2$ and bare NbSe$_2$, with the former exhibiting a reduced coherence peak height and a slightly larger gap size, as indicated by the red and black dashed lines. Upon increasing temperature to $T$ = 5 K [Fig. 2(d)], a shallow superconducting gap remains evident on both CrBr$_2$/NbSe$_2$ and bare NbSe$_2$, with nearly identical gap depths. Evolution of the superconducting gap spectra with perpendicular magnetic field ($\mathbf{B}_\perp$) demonstrates remarkably similar behavior for bare NbSe$_2$ and CrBr$_2$/NbSe$_2$ [Figs. 2(e) and 2(f)]. As $\mathbf{B}_\perp$ increases, the superconducting gap is gradually suppressed and becomes nearly indiscernible at $\mathbf{B}_\perp \sim$ 5 T, and the evolution of zero-energy conductance (ZEC) with $\mathbf{B}_\perp$ for both CrBr$_2$/NbSe$_2$ and bare NbSe$_2$ is displayed in Fig. 2(g). In both cases, the ZEC increases approximately linearly with $\mathbf{B}_\perp$, approaching unity at $\mathbf{B}_\perp \sim$ 5 T, which indicates the complete suppression of superconductivity at this field. These phenomena suggest similar superconducting state behavior for CrBr$_2$/NbSe$_2$ and bare NbSe$_2$.

As established, CrBr$_2$ is insulating, so there cannot be proximate superconductivity on it; instead, it serves as a vacuum tunneling barrier, the electrons from the STM tip directly tunnel through it to the underlying NbSe$_2$, detecting the electronic states of NbSe$_2$. Consequently, the superconducting gap spectra obtained from the CrBr$_2$ film directly reflect the intrinsic properties of the underlying NbSe$_2$ layer. Therefore, our results demonstrate that the superconducting state of NbSe$_2$ beneath CrBr$_2$ remains largely unaltered relative to bare NbSe$_2$, indicating minimal influence of CrBr$_2$ film on the superconductivity of NbSe$_2$. As shown in Fig. 1(g), the presence of CrBr$_2$ induces an upward band shift of $\sim$ 30 meV in the underlying NbSe$_2$, likely responsible for the minor superconducting gap variation. However, the variation in gap size is comparable to the energy resolution of our STM [43], thus falls beyond the scope of this discussion.

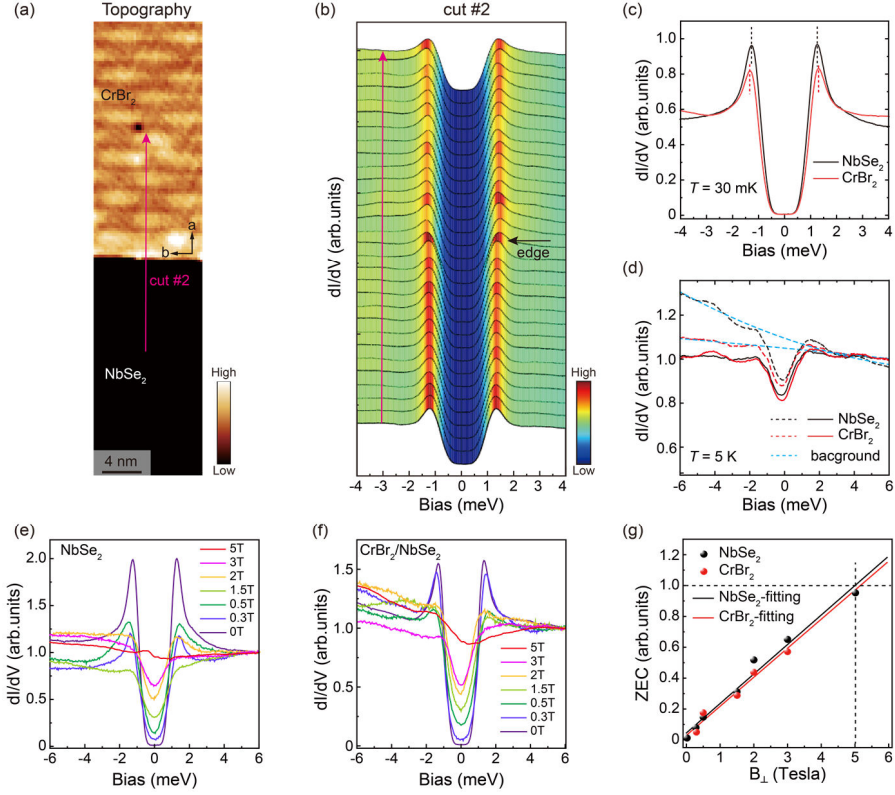

**Fig. 2. Superconducting properties of CrBr$_2$/NbSe$_2$ heterostructure**. (a) Topographic image of a selected sample region of CrBr$_2$/NbSe$_2$ heterostructure, including a single-unit-cell-height step. (b) Spatially resolved superconducting gap spectra acquired along the magenta line (cut #2) in panel (a). (c),(d) Averaged superconducting gap spectra on bare NbSe$_2$ and CrBr$_2$/NbSe$_2$, measured at $T \sim 30$ mK and 5 K, respectively. The red and black dashed lines in panel (c) denote the coherence peak positions. The black and red dashed lines in panel (d) show the raw data measured on bare NbSe$_2$ and CrBr$_2$/NbSe$_2$, while the black and red solid lines represent the fitted results by subtracting the polynomial backgrounds as indicated by the blue dashed curves. (e),(f) Evolution of normalized superconducting gaps as a function of $\mathbf{B}_\perp$, measured on bare NbSe$_2$ and CrBr$_2$/NbSe$_2$, respectively. (g) Evolution of ZEC as a function of $\mathbf{B}_\perp$, extracted from panels (e) and (f). Linear fitting is employed to determine the upper critical field, $\mathbf{H}_{c2}$. Measurement conditions: (a) $V_b = 2.5$ V, $I_t = 20$ pA; (b)-(d) $V_b = 6$ mV, $I_t = 100$ pA, $\Delta V = 0.2$ mV; (e),(f) $V_b = 6$ mV, $I_t = 100$-$200$ pA, $\Delta V = 0.1$-$0.2$ mV.

### C. Magnetic vortex of CrBr$_2$/NbSe$_2$ heterostructure

Figures 3(b)-3(d) present ZEC maps acquired within the same field of view (FOV) as Fig. 3(a) (outlined the CrBr$_2$ film by white lines), revealing the magnetic vortex distribution under varying $\mathbf{B}_\perp$. At $\mathbf{B}_\perp = 0.3$ T and 1 T [Figs. 3(b) and 3(c)], distinct magnetic vortices form a regular triangular lattice, showing no significant pinning by the CrBr$_2$ film or its boundary. Near the upper critical field ($\mathbf{B}_\perp = 5$ T $\approx H_{c2}^\perp$ of NbSe$_2$), superconductivity is completely suppressed and no magnetic vortices are observed [Fig. 3(d)]. Fitting the spatial evolution of ZEC between vortex cores with a two-dimensional Gaussian function yields the averaged nearest-neighbor vortex spacing (*d*) versus $\mathbf{B}_\perp$ [Fig. 3(e)], which closely matches the theoretical prediction for an Abrikosov lattice (red curve), $d = \left(\frac{4}{3}\right)^{\frac{1}{4}} \left(\frac{\Phi_0}{\mathbf{B}_\perp}\right)^{\frac{1}{2}}$, where $\Phi_0$ is the quantized flux. Exponential fitting of ZEC for individual vortices on CrBr$_2$/NbSe$_2$ and bare NbSe$_2$ [Fig. 3(f)], reveals similar coherence lengths ($\xi = 14.2 \pm 1$ nm and

12.5 ± 0.8 nm, respectively). The spatially resolved evolution of vortex bound states along the white arrows in Fig. 3(b) at $B_\perp$ = 0.3 T is shown for both CrBr$_2$/NbSe$_2$ and bare NbSe$_2$ in Figs. 3(g) and 3(h). In both cases, the bound states exhibit an "X"-shaped dispersion, precluding the existence of MZMs, which typically persists at zero energy over a long distance. These vortex characteristics are consistent with previous observations in bulk NbSe$_2$ [51,52], further demonstrating the weak influence of the magnetic CrBr$_2$ layer on NbSe$_2$.

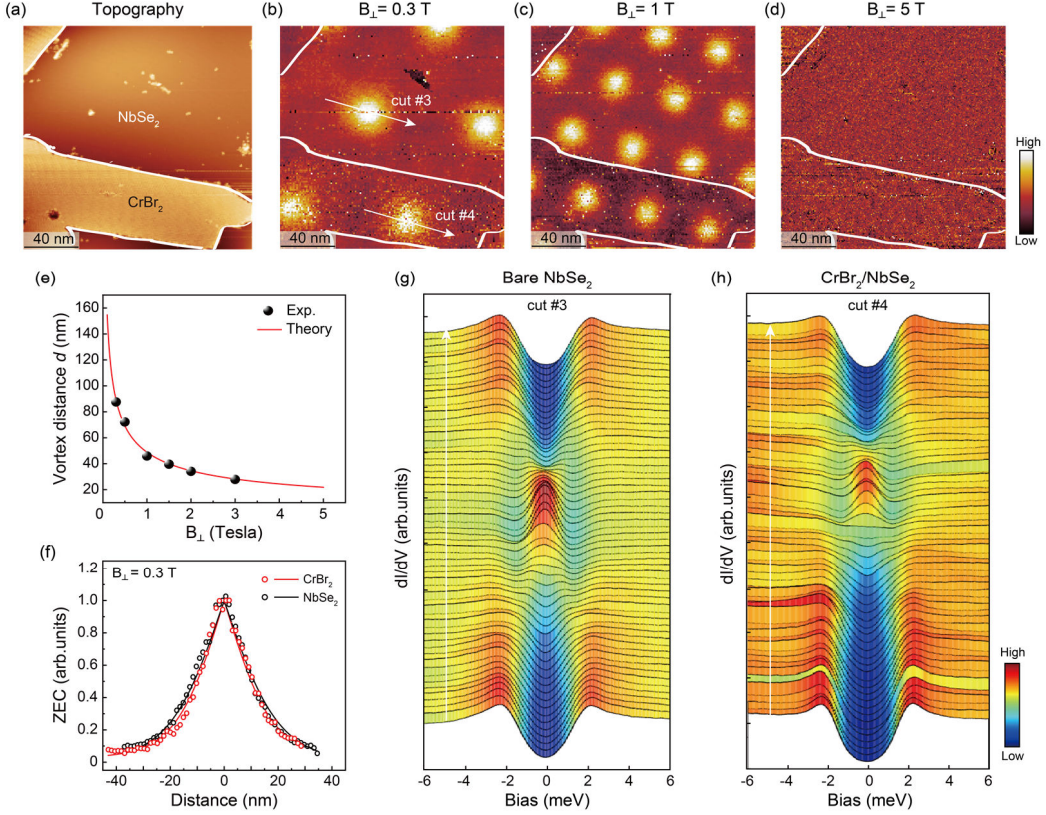

**Fig. 3**. **Magnetic vortices and vortex bound states of CrBr$_2$/NbSe$_2$ heterostructure**. (a) Topographic image of a selected sample region of CrBr$_2$/NbSe$_2$ heterostructure. (b)-(d) ZEC maps under various $B_\perp$, collected in the same FOV of panel (a). The CrBr$_2$ film is outlined by the white lines. (e) The averaged nearest-neighbor vortex spacing $d$ versus $B_\perp$, the red curve shows the theoretical prediction for an Abrikosov lattice. (f) Exponential fitting of ZEC for individual vortices on CrBr$_2$/NbSe$_2$ and bare NbSe$_2$ under $B_\perp$ = 0.3 T. (g),(h) Spatially resolved evolution of vortex bound states along the white arrows in panel (b) at $B_\perp$ = 0.3 T for both CrBr$_2$/NbSe$_2$ and bare NbSe$_2$. Measurement conditions: (a) $V_b$ = 6 mV, $I_t$ = 100 pA; (b)-(d) $V_b$ = 6 mV, $I_t$ = 100-200 pA, $\Delta V$ = 0.5-1 mV; (g),(h) $V_b$ = 6 mV, $I_t$ = 200 pA, $\Delta V$ = 0.1 mV.

### D. Edge states of CrBr$_2$ film

Theoretically, two-dimensional topological superconductivity supporting chiral Majorana modes can be realized in heterostructures combining a noncollinear magnet with a conventional s-wave superconductor [30-34]. Motivated by these predictions, we investigate the edge states at the boundary of CrBr$_2$/NbSe$_2$ heterostructures. Figures 4(a) and 4(b) show the typical topographic images of two distinct CrBr$_2$ island edges. Edge-I is a clean, unidirectional sharp step with few adsorbed clusters, while Edge-II exists near a domain boundary, exhibiting an irregular shape with numerous adsorbed clusters. Spatially resolved tunneling spectra acquired along both edges [Figs. 4(c) and 4(d)] reveal a widespread hard superconducting gap in clean regions. In contrast, adsorbed clusters exhibit suppressed superconducting gaps and obvious in-gap bound states. These in-gap states manifest diversely: either as an elevated gap bottom with suppressed coherent peaks [Fig. 4(e)], or a pair of energy-symmetric conductance peaks [Fig. 4(f)], or a zero-energy conductance peak [Fig. 4(g)]. Moreover, even at the same edge terminations, the in-gap states vary significantly, for examples, at point 5 on edge I and point 16 on edge II, as well as points 12-15 on edge II. This might be related to the variations in the origin of the clusters, as well as their adsorption position and arrangement form.

The presence of a hard superconducting gap and the direct correlation between in-gap states and adsorbed clusters are inconsistent with the expected behavior of topologically protected Majorana modes, which should form spatially continuous edge states. Furthermore, the insulating nature of the $CrBr_2$ film and the adsorbed clusters, combined with the highly spatially localized and diverse energy characteristics of the in-gap states, render the formation of Andreev bound states highly unlikely. Instead, these features closely resemble those observed in $CrBr_3$/$NbSe_2$ heterostructures, where in-gap states were identified as Yu–Shiba–Rusinov (YSR) states originating from local magnetic moments due to lattice reconstruction of step edge [47]. Here, the adsorbates could be molecular clusters or secondary phases of $CrBr_2$, $CrBr_3$, or other Cr-based compounds, which carry local magnetic moments, resulting in the formation of YSR states.

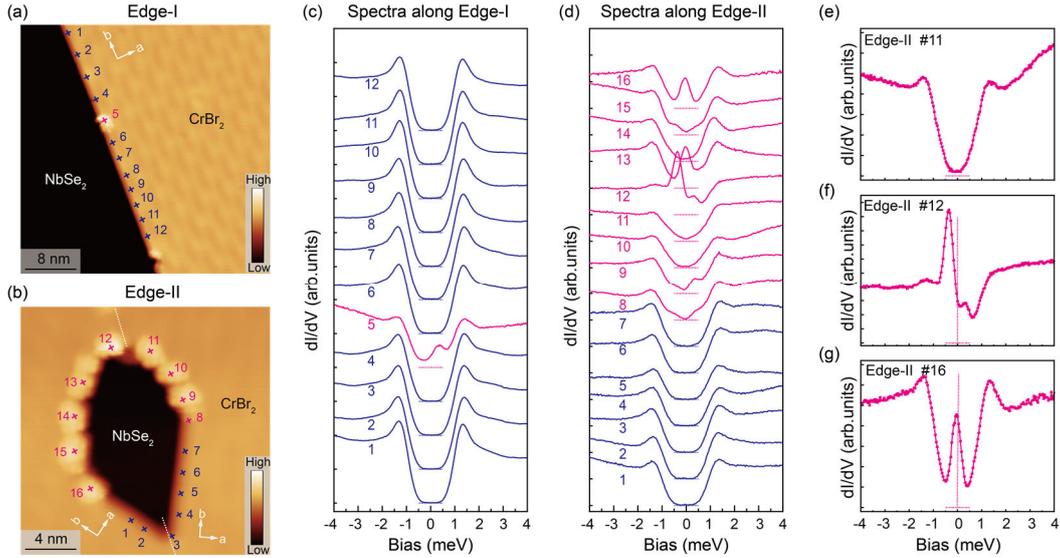

**Fig. 4. Influence of step edges on superconductivity**. (a),(b) Topographic images of two $CrBr_2$ island edges. In panel (b), the white dotted line indicates a twin boundary with different lattice orientations on its two sides. (c),(d) Superconducting gap spectra collected at the positions marked by numbered crosses along Edge-I and Edge-II in panels (a) and (b), respectively. The spectra are shifted vertically for clarity. (e)-(g) Three representative kinds of in-gap bound states collected along Edge-II. Horizontal dotted lines indicate the zero conductance for each spectrum. Measurement conditions: $V_b$ = 6 mV, $I_t$ = 100 pA, $\Delta V$ = 0.2 mV.

## IV. Discussion and conclusion

Overall, our study reveals multiple experimental signatures in $CrBr_2$/$NbSe_2$ heterostructures which point to weak interfacial coupling effects, preventing the system from entering the topologically nontrivial regime:

(i) The CrBr$_2$ film is insulating, preventing the superconducting proximity effect within the magnetic layer itself.

(ii) The superconducting properties of NbSe$_2$ covered by CrBr$_2$ film are strikingly similar to bare NbSe$_2$, and crucially, no topologically nontrivial edge states are realized. These results suggest that the helimagnetism (and potential ferroelectricity) of CrBr$_2$ exerts only a very weak influence on the underlying superconductivity of NbSe$_2$.

(iii) The direct signature of strong interfacial coupling — band splitting or Shiba bands residing within the superconducting gap — is notably absent in CrBr$_2$/NbSe$_2$ heterostructures.

This situation parallels our previous findings in heterostructures composed of ferromagnetic insulator CrBr$_3$ and superconductor NbSe$_2$ [47], as well as in several other magnet/superconductor van der Waals (vdW) heterostructures, such as MnTe/NbSe$_2$ [48], CrCl$_3$/NbSe$_2$ [50], and MnSe/NbSe$_2$ [53]. A critical question arises: why does the theoretically proposed model for realizing topological superconductivity via the coupling of magnetism and superconductivity fail to materialize in those heterostructures? To gain deeper insights, we compare the detailed properties of these magnet/superconductor heterostructures.

Firstly, the superconducting layers are the same, all being NbSe$_2$; however, the materials in the magnetic layers differ. CrBr$_2$ is a helimagnet, adopting the typical crystal structure of 1T-phase transition metal dichalcogenides (TMDs) but with significant Jahn-Teller distortion [Fig. 1(a)]. CrBr$_3$ and CrCl$_3$ adopt the rhombohedral BiI$_3$ structure [54], both are ferromagnets but the spin orientations differ in the monolayer form [55,56]. Monolayer MnSe and MnTe possess an unusual atomic structure similar to the layered CuI, and DFT calculations suggest an antiferromagnetic order [53,57]. These materials with completely different crystal structures and magnetic structures, when coupled with NbSe$_2$ superconductor, result in similar weak interfacial coupling effects, indicating that the structural and magnetic properties are not the primary factors.

Secondly, these magnetic materials possess distinct electronic structures, but all of them are insulators with large band gaps, which maintain when epitaxially grown on NbSe$_2$ [47,48,50]. Therefore, superconductivity cannot be introduced into the magnetic layers; instead, the interfacial coupling effect can only be manifested in the influence of magnetism on superconductivity of the underlying NbSe$_2$ layer. We attempt to estimate the magnetic exchange coupling strength $J$ that can act on the superconducting layer. In theoretical models [30-34], it is possible for the magnet/superconductor system to enter the topologically nontrivial regime only when $J$ is larger than the superconducting gap $\Delta$. Based on the empirical law that the direct exchange coupling strength $J$ decays exponentially with the spatial separation between atomic sites [58], i.e. $J \sim J_0 \exp(-2kd)$, here $J_0$, $k$, and $d$ represent the in-plane nearest-neighbor magnetic exchange coupling strength in the magnetic layer, the decay constant and the interlayer distance, respectively. From neutron scattering studies, we know that $|J_0|$ is approximately 1-1.5 meV for the nearest-neighbor Cr-Cr pairs in CrX$_2$ and CrX$_3$ systems [27,28,59-61]; the typical value of $k$ is ~ 0.5-1.5 Å$^{-1}$, and $d$ is ~ 4-6 Å for the studied heterostructures here. Therefore, the exchange coupling between Cr and Nb atoms in the topmost layer of NbSe$_2$ is $J_{\text{Cr-Nb}} \ll 0.1$ meV, much smaller than the superconducting gap size $\Delta \sim 1.2$ meV for NbSe$_2$. Obviously, the topological nontrivial criterion $J > \Delta$ is not satisfied in these systems, which successfully explains why the *s*-wave superconducting characteristics of NbSe$_2$ are still maintained at the magnet/superconductor interface. In contrast, in heterostructures such as

Fe/Pb that are not fabricated by vdW epitaxy, the *d* value is very small while *J* is much larger, making it easier to enter the topologically nontrivial regime [11,13].

Based on above analysis, we know that enhancing the interfacial coupling effect is paramount for exploring unconventional and topological superconductivity in magnet/superconductor vdW heterostructures. In future studies, 2D magnetic metals or magnetic semiconductors with smaller band gaps can be considered. The former are metals and the latter can probably enter weak metallic phases through interfacial charge transfer, which promote the occurrence of superconducting proximity effects. Moreover, the presence of charge carriers can also introduce other types of magnetic interactions to enhance the interfacial coupling.

## Acknowledgments

This work is supported by the National Natural Science Foundation of China (Grants No. 12374140, No. 12074363, No. 12494593), the Innovation Program for Quantum Science and Technology (Grant No. 2021ZD0302803), the National Key R&D Program of the MOST of China (Grants No. 2023YFA1406304), the New Cornerstone Science Foundation.

## Data availability

The main data supporting the findings of this study are available within the article. All the raw data generated in this study are available from the corresponding author upon reasonable request.